\documentclass[conference]{IEEEtran}
\IEEEoverridecommandlockouts
\usepackage{cite}
\usepackage{amsmath,amssymb,amsfonts}
\usepackage{algorithmic}
\usepackage{graphicx}
\usepackage{textcomp}
\usepackage{xcolor}
\usepackage{graphicx}
\usepackage{fancyhdr}

\usepackage{times}
\usepackage{epsfig}

\usepackage[breaklinks=true,bookmarks=false]{hyperref}

\def\BibTeX{{\rm B\kern-.05em{\sc i\kern-.025em b}\kern-.08em
    T\kern-.1667em\lower.7ex\hbox{E}\kern-.125emX}}
\begin{document}

\title{State of Security and Privacy Practices of Top Websites in the East African Community (EAC)\\
}

\author{\IEEEauthorblockN{Abdirahman Mohamed}
\IEEEauthorblockA{\textit{Carnegie Mellon University} \\
\textit{College of Engineering}\\
Kigali, Rwanda \\
mabdirah@andrew.cmu.edu}
\and
\IEEEauthorblockN{Christopher Dare}
\IEEEauthorblockA{\textit{Carnegie Mellon University} \\
\textit{College of Engineering}\\
Kigali, Rwanda \\
cdare@andrew.cmu.edu}
\and
\IEEEauthorblockN{Ayobami Esther Olanrewaju}
\IEEEauthorblockA{\textit{Carnegie Mellon University} \\
\textit{College of Engineering}\\
Kigali, Rwanda \\
aolanrew@andrew.cmu.edu}
\and
\IEEEauthorblockN{Mercyleen Tanui}
\IEEEauthorblockA{\textit{Carnegie Mellon University} \\
\textit{College of Engineering}\\
Kigali, Rwanda \\
mtanui@andrew.cmu.edu}
\and
\IEEEauthorblockN{Fonyuy Boris Lami}
\IEEEauthorblockA{\textit{Carnegie Mellon University} \\
\textit{College of Engineering}\\
Kigali, Rwanda \\
fborisla@andrew.cmu.edu }
}

\maketitle

\begin{abstract}
Growth in technology has resulted in the large-scale collection and processing of Personally Identifiable Information (PII) by organizations that run digital services such as websites, which led to the emergence of new legislation to regulate PII collection and processing by organizations. Subsequently, several African countries have recently started enacting new data protection regulations due to recent technological innovations. However, there is little information about the security and privacy practices of top websites serving content to EAC citizens.  We, therefore, analyze the website operator's patterns in terms of third-party tracking, security of data transmission, cookie information, and privacy policies for 169 top EAC website operators using WebXray, OpenSSL, and Alexa top websites API. Our results show that only 75\% of the analyzed websites have a privacy policy in place. Out of this, only 16\% of the third-party tracking companies that track users on a particular website are disclosed in the site’s privacy policy statements which means that users don’t have a way of knowing which third parties collect data about them when they visit a website. Such privacy policies take time to read and are difficult to understand; on average, it takes a college graduate to comprehend the policy and a user spends 12 minutes to read the policy. Additionally, most third-party tracking on EAC websites are related to advertisement and belong to companies outside the EAC. This means that EAC lawmakers need to enact suitable laws to ensure that people's privacy is protected as the rate of technology adoption continues to increase.

\end{abstract}


\section{Introduction}
Privacy, as Nissenbaum puts it, "is not one-size-fits-all; it varies by context." Privacy has many ways one can view it. For example, it can be taken as a concept (regulating access to and control of one's data between self and others) or as a right (right to delete one's data, reject surveillance, or update information collected about oneself, etc.) [1]. The people's perception of privacy is not often seen the same way in the eyes of lawmakers. It is also important to note that the advancement of technology, especially in the last 20 years, has presented new threats and opportunities that impact the quality of user privacy, which is evident from the interactions between website operators, users, and third parties. 

Another crucial aspect concerns the security of website data transmission as some political regions, like the European Union (EU), have implemented sound regulations that influenced how website operators conduct business in such territories. Other countries like the United States of America (USA) and China are beginning to catch up by developing data protection regulations that reasonably protect and preserve user privacy in a manner that still fosters economic development with technology [2,3]. Unfortunately, many African countries are yet to catch up with the trend of enacting adequate legislation that will advocate for and enforce their citizens' data protection. Out of the six countries in the EAC, only Kenya and Uganda have enacted a data privacy law to protect their people. However, it is worth noting that Rwanda is currently in the process of passing its privacy laws. As we have observed with many other countries like the EU and USA, knowledge and research provide a secure foundation for decision and policy making on data protection. This lack of adequate data privacy policies can open up Africa to privacy rights violation. Technology advancement has allowed the internet, mobile phones, and computers to be of our daily use, making data more portable. 

From works of literature, very little research has been done on website operator’s security and privacy practices such as third party tracking, privacy policies effectiveness, and the security of data transmission in the context of popular websites in the EAC. We utilized Alexa’s top websites API to identify top EAC websites (N=169) across the six member states and used the WebXray tool to collect third party tracking, cookie, and privacy policies data. We also used OpenSSL to collect data related to security of data transmission on websites. Our findings reveal that top uses of third party cookies on websites are for marketing, audience measurement, and hosting with only 16\% of third parties that track website users disclosed through the privacy policies. Also, majority of third-party tracking on EAC are linked to Google followed by other Analytics and Advertisement (A\&A) companies and are all located in the US.

We also found that privacy policies of top EAC websites are written at a grade level of 29 on the Flesch Reading Ease (FRE) scale which means that they are very difficult to read and that an education level of a college graduate is needed to adequately comprehend these policies. Interestingly, 1.2\% of the total websites have expired TLS certificates while 35.5\% use an insecure version of TLS (TLSv1.2).

The following subsections address the definition of terms, explain the topic areas, problems to be addressed and the context of our research.

\subsection{\textbf{Definition of Terms}}
\textbf{Privacy}: This is the ability to keep information about someone secret. And can be likened to controlling access to data which is shared with someone else.

\textbf{Privacy Policy} : A privacy policy is a legal document that specifies how a company handles user’s data, be it their customer or a website user. This relates to collecting, organizing, storing, manipulating, analyzing, and processing such data.It is also known as policy notice or policy statement.

\textbf{Privacy Compliance}: Privacy compliance is the ability to conform to the laid down rules regarding the protection of the personally identifiable information of users.

\textbf{Third-Party Tracking}: This is the practice by which an entity (the tracker), other than the website directly visited by the user, tracks or assists in tracking the user’s visit to the site [4].

\textbf{Cookies}: These are small files stored on a user’s computer by websites for the purpose of user identification and personalisation of the user browser experience [5].

\textbf{HTTP Requests}: This is a request sent by a client such as a web browser to web server to fetch a resource hosted on the web server [6].

\textbf{EAC}: Stands for East African Community which is a regional intergovernmental organisation comprising six partner states, namely Burundi, Kenya, Rwanda, South Sudan, Tanzania and Uganda. It has its headquarters in Arusha, Tanzania [7].

\textbf{Data Protection Regulation/Law}: This is a legislation enacted by a country/regional block with the aim of protecting the process of collecting, storing, and processing of PII of users or citizens [8].

\textbf{PII}: PII is an acronym for Personal Identifiable Information. This is information which can be used to uniquely identify an individual and includes information such as social security number, address, phone number, etc.

\subsection{Visibility of Website Operator Activity}

It is not always evident to users how website operators handle user data or even those they share with the collected data. We try to solve this by identifying and analyzing websites’ data sharing practices with third parties. The result of this might, in turn, inform the formulation of adequate and realistic data protection and privacy laws. 

\subsection{Effectiveness of Privacy Policies}

Privacy policies on websites communicate to users crucial information such as what kind of data a website collects and processes, the use of such data and who they share it with. A good understanding of this will help users make the right decisions on interacting with these websites. However, most often, the privacy policies on these websites are not easily accessible and challenging to comprehend [9]. Therefore, we seek to evaluate the privacy policies' readability for websites under review to empower policymakers and users alike. 

\subsection{Development of Data Protection Legislature}
With the proliferation of information technologies and the rising use of social media, coupled with the ease of data transfer, there is a need for countries in the EAC  to implement policies that will protect their citizens from data privacy violations. No doubt technologies are promising as they can affect different facets of an economy and significantly improve an economy. Improper handling of the data involved can leave a country vulnerable.

Out of the six countries in the EAC, only Kenya and Uganda have enacted a data privacy law to protect their people. However, it is worth noting that Rwanda is currently in the process of passing its privacy laws. While Burundi, Tanzania and South Sudan are yet to enact any data privacy policy. This knowledge about the EAC leaves us with the belief that there are attempts by these countries to formulate policies and regulations that mimic the General Data Protection Regulation (GDPR); however, a lot of African countries seem reluctant in developing such policies [10]. Considering that data protection is moving from just a might-have for business to a must-have as it tends to protect the fundamental human right to privacy [10]. The provision and implementation of these laws will help develop the countries through new technologies. It will also ensure citizens' rights are protected.

\subsection{Research Question}
We intend to expose the behaviors and practices relating to privacy and security of the top  websites in EAC member states. Specifically, we intend to: (i) Investigate key and accessible security elements of websites and determine the compliance and security of websites such as TLS Certificates. (ii) Analyze and expose the third-party activity on websites users visit (based on request and cookie data). (iii) Evaluate privacy policies to examine if they are present or not, their length and readability.
The impact of our research would be more on the protection of personal user data and raising awareness on the rights of individuals when it comes to privacy. The audience of our study includes but not limited to the following: Lawmakers of countries within the EAC, researchers interested in strengthening data privacy in Africa, and for Africans, website users seeking to understand the privacy and security practices of websites they use, website operators interested in data privacy regulations, and other stakeholders interested in understanding security and privacy practices of online service providers in EAC. 
\subsection{Scope}
Our research exposes top-visited websites' privacy and security practices in the six EAC member countries (Kenya, Uganda, Tanzania, Rwanda, Burundi, and South Sudan). The research does this by examining security and privacy parameters that might indicate good/lousy security/privacy practices exhibited by websites. We intend to investigate data sharing through frontend (browser based) HTTP requests only and not server side data sharing. These parameters include the level of sharing of user data with third parties by looking at cookie data and HTTP requests, the readability and of privacy policies posted on the websites, and basic security controls such as Transport Layer Security (TLS) encryption on the websites. The specifics of the parameters, as mentioned earlier, are discussed in detail under the methodology section. Anything outside this is not within the scope of our research and, as such, not covered.
\subsection{Assumptions}
Given that some of the EAC member states are yet to enact robust data protection and privacy legislation fully, we assume that some of the website operators' online service providers in these countries have lapses in how they handle the personal information of their users. We are also assuming that the implementation features of some websites, such as the utilization of third-party libraries, increase their attack surface, leading to a breach of personal information. We believe that the content and data transmission activities between websites, users, and third parties are the same across the EAC geopolitical region. We assume that crawling data obtained while in Rwanda generalizes to the EAC region. Lastly, given the reliance of today's websites on third-party plugins such as Google AdSense, we assume that such website operators share user data with these third parties.

\section{Literature Review}
The majority of the EAC member states either have stand-alone data protection regulations (in the draft/enacted state) or have addressed this in some form within their constitution. According to the United Nations Conference on Trade and Development (UNCTAD), Kenya enacted a Data Privacy and Protection legislation in 2020. Rwanda, Uganda, and Tanzania have draft legislation in place. Burundi and South Sudan do not have a data protection legislature [11]. It is worth noting that only Rwanda has signed the African Union's Malabo Convention on cybersecurity and data protection. The need to enact data protection and privacy legislation can partly be attributed to the increasing number of mobile subscribers in the EAC member states. GSMA reported in 2019 that the mobile subscriber penetration rate in EAC countries was almost half the population (49\%) and has plunged from 46\% in 2018 [12]. 

Little is known about popular online service providers' actual privacy and security practices in the EAC because prior work mainly focused on countries/regions outside the EAC. Those that narrowed their research to our area of interest mostly covered a small subset of online service providers such as e-government websites using techniques such as surveys and interviews. Therefore, it is crucial to validate the assumptions upon which current data privacy legislation in EAC has been designed by exposing popular websites' security and privacy practices in EAC. This way, users become more aware, which will, in turn, help them make informed and privacy-conscious decisions on how they interact with these websites. Consequently, we hope to expose the practices and behaviors of website operators that might violate the requirements of the respective countries' data protection legislation (prior work in other parts of the world has shown that there is reluctance by websites in adhering to legal and ethical principles [13]). This study would further help lawmakers understand the level of compliance of website operators with the enacted data protection and privacy legislation.

Consequently, these insights can yield privacy compliance by website operators, resulting in better patronage and business. When businesses protect the customer's data, customer's trust in the business increases and their reputation/brand improves [14]. It's therefore essential to formulate regulations that are not too tight and not too loose to ensure that business is not lost while protecting the privacy of citizens in the EAC.

The following sections present themes covered in prior related work.

\subsection{Privacy and Security Controls in E-government Sites}
The introduction of e-government services in most African countries led to increased efficiency, convenience, and accessibility of government services. However, there exist insufficient security and privacy controls needed to protect the processing of Personally Identifiable Information (PII) on e-government websites. Several studies have looked into the privacy and security practices of e-government sites mainly in West Africa with only few studies done in the context of East Africa. A study of privacy practices of e-government websites in Ghana, Nigeria, Sierra Leone, Gambia, and Liberia against ISO 29100 privacy principles and the ECOWAS Regional Data Protection Act revealed the existence of limited controls in protecting the PII of the respective countries’ citizens and complying with the above privacy principles/acts [15]. A separate study that examined PII processing by Nigerian consulates in 8 countries revealed the existence of ineffective controls in the protection of PII of Nigerian citizens living abroad [16]. Mutimukwe et al. evaluated compliance of three leading e-government service providers' in Rwanda with international privacy principles and found that the providers partially comply with the principles [17]. However, [15, 1] examined limited privacy and security parameters when evaluating compliance while [17] relied on survey and interview data to draw their conclusions. Our approach advances the aforementioned studies by evaluating websites' actual privacy practices using web scraping and data mining techniques.

\subsection{Privacy Concerns of Online Services Users in EAC}
Users in the EAC are concerned about the privacy of their personal data and are mindful of who they share such information with. For example, a study of privacy concerns of internet users revealed that East Africans are more concerned about their online privacy compared to the U.S. internet users. However, this result might not necessarily correspond to their actual behavior with respect to measures taken by East African users to protect their online privacy since the same study found that users in the US are more familiar with methods of covering their tracks online. This might be due to the Snowden revelations in 2013. The same study found that East Africans were more likely to report negative experiences related to social engineering attacks and online fraud [18]. This might be as a result of the high uptake of mobile and internet banking solutions by users in EAC in the recent past which makes them valuable targets for threat actors perpetrating fraud. It is for this reason that user awareness and formulation of adequate cyber and privacy legislation becomes paramount [19]. Additionally, users in Rwanda are increasingly concerned about the privacy implications of sharing their personal information with even some refusing to share their personal information with e-government websites [20]. It is therefore important to expose the security and privacy practices such as third party activities of websites in EAC so that users can make informed and privacy-conscious decisions on how they should interact with these websites.
\subsection{Third-Party Activity: Processing (Outsourcing) and Tracking}
Who else is seeing information about me when I visit this website? One of the significant privacy themes concerns the disclosure of user data to third parties. Consequently, researchers have tried uncovering 3rd party activity from interactions between users and websites they visit. We can mainly view 3rd party activity from 3rd party tracking and 3rd party data processing. Past evaluations of 3rd party tracking have shown that monitoring typically occurs through scripts (JavaScript), HTTP requests to 3rd parties, browser fingerprinting, and storing cookies on the users' devices accessible by third parties. Findings show that 3rd parties resolve to a few sets of prominent corporations, revealing that few companies own vast troves of user data. Notable amongst these companies are Google, Facebook, DoubleClick, and Akamai [21, 22]. Website operators in Africa also sometimes share user data with 3rd parties for processing. For instance, a privacy compliance study of e-government websites in five West African Anglophone countries revealed that 80\% of the countries outsourced the processing of their citizen's PII to foreign countries raising concerns of misuse of such information in cases such as identity theft [15].
The use of cookie synchronization protocol on websites also impacts users’ privacy by leaking assigned userIDs and sharing them with third parties which makes it easier in reconstructing user's data from their browsing history. From this, we learn that user data is the primary input of digital advertising and web companies invest a lot in elaborate tracking mechanisms to acquire user data that can sell to data markets and advertisers [23].
In addition to the above, the likelihood of a user receiving notice of the third-parties receiving their data by reading a site’s privacy policy is remarkably low yet the total time needed to read all applicable policies for a given site is 84.7 minutes on average [9]. Therefore, to conduct effective targeted investigations and develop sound data protection regulations, EAC regulators need to understand how website operators share citizens' data with 3rd parties. Tracing and investigating such activities can also help regulators evaluate their instituted data protection regulations' effectiveness and enforcement.

\subsection{Adequacy of Data Privacy and Protect Legislation in Africa}
There are significant concerns about data privacy in Africa, which can be attributed to Africa's low adoption of state-of-the-art digital techniques and technologies (such as artificial intelligence and big data) to track website operators' compliance. The ease of data migration to a third party to carry out a customized service also raises concerns about whether to ask if the user's consent was sought before carrying out these tasks. Furthermore, first-world countries have access to incredible technologies, which can further endanger Africa's privacy. If other governments can access African citizens' data without citizens' consent, how safe is their data? To better address these concerns, a synergy between academia and government, followed by the government enacting laws based on academia's research and following it to completion, will improve Africa's data privacy practices [10].
The rate at which quality data privacy and protection legislation are enacted is slower when compared to the technology growth rate. This issue has been augmented by the lack of widely accepted, comprehensive and overarching regulations that can be used by all African Countries. The AU came up with such an initiative, but it is yet to be adopted and accepted by the majority of the member states. Additionally, there are very few documented cases of people coming to court seeking redress for data privacy violations making it difficult to gauge the regulations’ reach. This can partly be attributed to user unawareness and lack of unified data privacy laws [24].

\subsection{Identification of PII}
While websites should provide detailed descriptions about the information, they will collect from a web user and the data's use, some websites do not disclose 100\% what information they are collecting and its extent [16]. The PII, an acronym for Personally Identifiable Information, provides a means to identify an individual uniquely. This information could range from name, address, social security number, date of birth, passport numbers to credit card numbers. We can also categorize PII as SPII (Sensitive Personal Identifiable Information); this is information that, if lost, compromised, disclosed, could cause harm, embarrassment, unfairness to an individual. Examples of such information include medical information, sexual orientation, ethnic or religious identity. Misuse of an individual's personal information as they transact on the Web involves the unauthorized acquisition, disclosure and use of personal information. Since 1973, numerous guidelines, principles, conventions, directives, acts, and reports in the European Union (EU) and the United States have implemented a framework that should be followed by organizations that collect PII, both offline and online [16].
Additionally, for the websites that disclose this information, the user should decide whether these policies are satisfactory and whether he or she will provide his/her PII to this organization; This is the idea behind posting privacy policies on a website. Consumers must read these policies and make informed and rational choices. Privacy, trust, and security are important dimensions of website quality in these studies. Despite their significance, this area still needs extensive research in the context of EAC to evaluate websites and analyze their methods for treating private information according to their privacy policies [16]. 

We realize from this prior work, that the topic area of our study is rather a grey area with little research done in the context of the EAC. This forms the basis and motivation for our research.

\section{Methodology}
Drawing inspiration from Libert et al., we attempt to reveal the security and privacy practices of top websites in the EAC by analyzing crucial components of website data: security of data transmission, cookies, third party requests and privacy policies. Libert et. al have shown that the first three components reveal a lot concerning the website operator’s practices [9]. Figure 1 describes our approach to analyzing the data which is further explained below.

\begin{figure*}
    \centering
    \includegraphics[width=\textwidth]{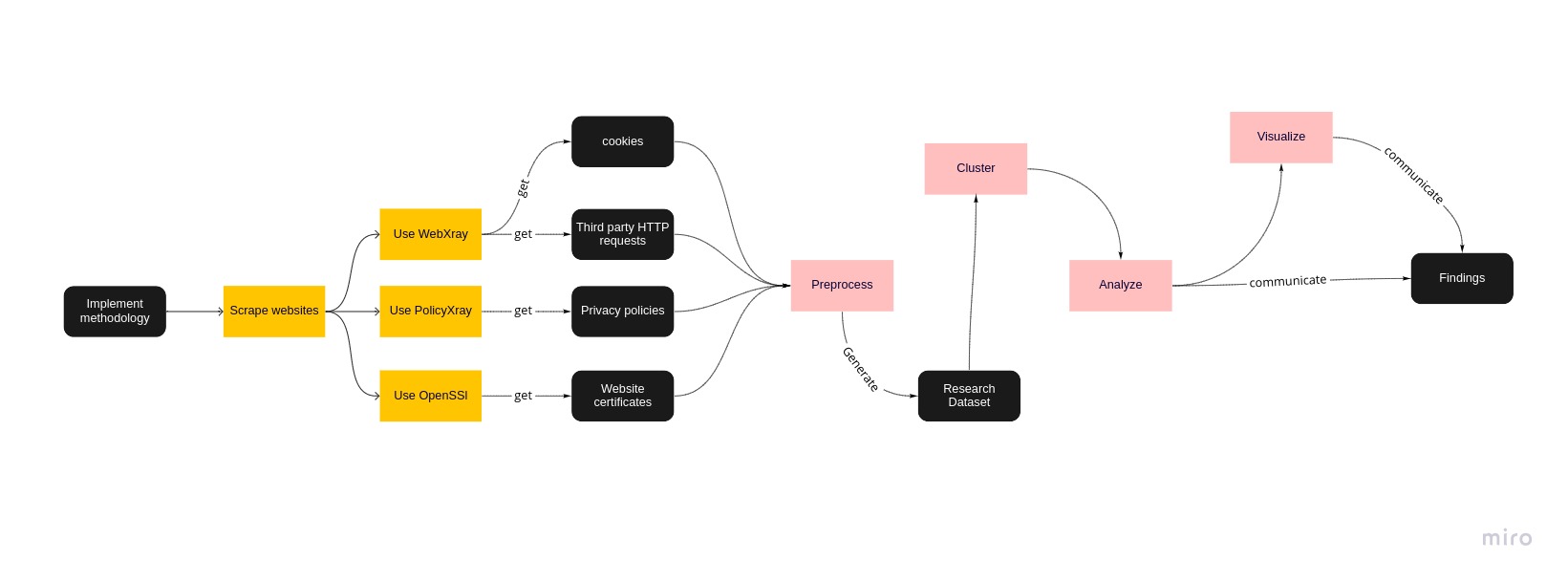}
    \caption{Research Methodology}
    \label{fig:my_label}
\end{figure*}

We create our own dataset of website metadata from top 169 websites used by residents of the EAC region. The ranked list of websites was retrieved from Alexa [25]. The website data was mined through WebXray (a tool for analyzing website HTTP traffic, page and cookie content, extracting legal policies, and identifying the third party companies which collect user data [26]) and OpenSSL from top websites within the EAC. Data collected with WebXray adequately represents the experiences of most desktop web users.

Using Alexa’s top websites API on AWS, we retrieved the top 50 websites each for Kenya, Uganda and Tanzania[27][28][29]. This data set comprised local and international websites. Since Alexa’s API did not have a dataset of top websites for Burundi, Rwanda and South Sudan, we curated our dataset of websites from the internet which, by empirical and qualitative analysis, would be considered as websites patronized in those 3 countries. Based on similarities in website patronage by users in our Alexa dataset, we reckon that websites that may not have been identified in our qualitative analysis as top websites for Rwanda, Burundi and South Sudan, would likely be found in the Alexa dataset. Our empirical analysis yields a dataset of websites patronized in Rwanda and Burundi[30][31][32]. We could not find a credible means of estimating top/highly patronized in South Sudan. However as mentioned, our dataset contains websites that are highly likely to be patronized in South Sudan. One such example is youtube.com. Concatenating these 2 data sources resulted in our dataset of 169 unique websites.

\subsection{HTTP requests and cookie data}
The beauty of interactions between website operators and third parties is that they take place via HTTP requests. Consequently, we can identify and track interactions between a website operator and 3rd parties using WebXray.  By monitoring information such as request url information, TLS encryption, we can answer the following questions: (i) Which third parties are involved in interactions between a user and the website operator? (ii) What categories do these third parties belong to? (iii) Are these connections secure?.

The answers to these questions will help lawmakers evaluate website operator’s practices and understand how best to control their activities that might lead to privacy violation. Website users can also be informed of the 3rd parties who potentially have access to their data and can query a website operator to confirm the relationship between a 3rd party and their (users’) data. 

\subsection{Privacy policies}
Privacy policies are the means through which website operators indicate to users, the terms of agreement of websites which concern their privacy by stipulating how they collect, store, and process their data. We used WebXray’s PolicyXray module to collect website privacy policies. Consequently, analyzing privacy policies will also help us answer the following questions: (i) Does the website have a privacy policy/statement?, (ii) How long is the website's privacy policy/statement?, (iii) How easy is it for a user to read and understand the policy?, (iv) Are third parties that track users on a particular website disclosed through the privacy policy?.

The ease of reading a privacy policy was based on the Flesch Reading Ease (FRE) scores which range from 0 to 100 with lower scores signifying that a piece of text is more challenging to read [33]. The answers to the above questions will provide privacy regulatory bodies within the EAC with the insights for enforcing privacy regulations that are inline with the current state of affairs as far as website’s privacy policies are concerned.

\subsection{Security of Data Transmission}
Without access to a website operator’s code and technology infrastructure - as well as that of third parties, it is rather impossible to holistically evaluate the security of a website. By analyzing the TLS and certification metadata, we can answer the following questions: (i)Is the website using TLS Certificate or not?, (ii) Which TLS version?, (iii) Has the certificate  expired?

The answers to these questions will reveal to lawmakers and regulatory bodies (to some extent) the state of security of data transmission between users, website operators and third parties. Website operators may also use these insights to take actions such as enforcing TLS security on their websites or cutting third parties who do not have such levels of security.

\section{Results and Discussion}
\subsection{Third Party Tracking and Cookie Data}
Although users may think they are interacting with one entity from the address on the browser, most sites include code from other parties (third parties) to collect information about the user which users are typically unaware of [34]. We found that 3rd party interactions are much more compared to 1st party for the websites under study. The total number of websites under review were 169 but only managed to collect third party and cookie information of 129.
The number of third party cookies against the total cookies is $54.35\%$ higher than first party cookies! Additionally, 3rd party requests and responses on websites are leading by 58.85\% and 57.46\% respectively.
\begin{table}
    \centering
    \begin{tabular}{|c|c|c|c|}
    \hline
        \textbf{Owner} & \textbf{Request Count} & \textbf{Percentage} & \textbf{Industry}  \\
    \hline
    Alphabet (Google) &  181 & 36.20\% & Marketing \\ \hline
    Facebook & 40 & 8.00\% & Social media \\ \hline 
    Twitter & 16 & 3.20\% & Social media \\ \hline
    Amazon & 14 & 2.80\% & Ecommerce \\ \hline
   Others & 249 & 0.50\% & Mixed \\ \hline
    \end{tabular}
    \caption{Top Third-Party requests by Ownership}
    \label{tab:top_third_party_requests_by_ownership}
\end{table}\\
From the top 5 third party requests, we see from Table \ref{tab:top_third_party_requests_by_ownership} above that Google, through its parent company Alphabet, takes a huge chunk of incoming requests from websites users visit. These requests were particularly sent for advertising and tracking purpose (i.e. Adsense and Google analytics). We found it interesting that 36.2\% of third-party requests are routed to Google - but also that a substantial quantity of user data of EAC residents is shared with Amazon.
\begin{table}[]
    \centering
    \begin{tabular}{|c|c|c|c|}
    \hline
           & \textbf{Percent } & \textbf{Percentage} & \textbf{Percentage}  \\
            \textbf{Use case}    & \textbf{ crawls} & \textbf{of use w } & \textbf{of use}  \\ 
                                & \textbf{ with use} & \textbf{cookie} & \textbf{ssl}  \\ \hline

        Marketing &
        81.4 &
        1.23 &
        99.96 \\ \hline
        Hosting &
        75.97 &
        3.56 & 
        99.64 \\ \hline
        Audience measurement &
        68.22 &
        0.59 &
        99.73 \\ \hline
        Tag manager &
        58.91 &
        0 &
        100 \\ \hline
        Others &
        28.42 &
        37.65 &
        99.82 \\ \hline
    \end{tabular}
    \caption{Top five third-parties by use cases}
    \label{tab:top_five_third-parties_by_case_cases}
\end{table}
\\

Out of the total crawls made, the top five use cases of 3rd party cookies on EAC websites are: marketing which amounts to 81. 4\%, followed by hosting at 75.97\% and lastly for audience measurement at 68.22\%. From this information, we can infer that the third parties could be Advertising and Analytics (A\&A) companies whose core business is advertising.\\
From the table, 99.73\% of third-party cookies used for audience measurement were utilizing SSL. Meaning the remaining 0.27\% of the cookies used unsecure links. This still raises the possibility of data being intercepted owing to the insecure transmission. This interception can reveal user’s privacy data that may be present in the cookies and allow for third-party identification and tracking.\\
From the results we got, we were able to identify the leading country where these third-party companies are based to be the US.\\
\subsection{Privacy Policies}
Out of the 169 websites we identified, we successfully extracted privacy policies for 127 websites which represent 75\% of the total. Some of the reasons for not successfully extracting a privacy policy from a website might be that the website does not have a policy at all, the link to the policy uses uncommon phrases, or the structure of the page makes it difficult to extract the policy. Additionally, we found that policies for top websites in the EAC are written at a grade level of 29 on the Flesch Reading Ease (FRE) scale which means that they are very difficult to read and that an education level of a college graduate is needed to adequately comprehend these policies.\\
Moreover, we found that the analyzed policies have an average word count of 2981. Based on McDonald and Cranor’s method which estimated the average reading rate of privacy policies to be 250 words per minute [35], we infer that it takes East Africans an average of 12 minutes to read a privacy policy on a website. When it comes to the disclosure of third parties who track the website’s users, we found that only 16\% of such third parties are listed in the privacy policy indicating that users have no way of learning about third parties who collect and process data about them on a particular website.

\begin{table}
    \centering
    \begin{tabular}{|c|c|c|c|}
    \hline
        \textbf{Privacy Policy Property} & \textbf{Quantity}  \\
    \hline
    N &  127 \\ \hline
    Average Word Count & 2981 \\ \hline 
    Average FRE Score & 29 \\ \hline
    \% of 3rd Parties Disclosed & 16 \\ \hline
   Average Read Time Per Policy & 12 minutes \\ \hline
    \end{tabular}
    \caption{Privacy Policies Summary}
    \label{tab:privacy_policies_summary}
\end{table}

\subsection{Security of Data Transmission}
We observed that over 63.3\% of the top websites in the EAC use TLSv1.3, 35.5\% use TLS v1.2, and 2.2\% do not use any form of transport layer encryption. TLSv1.3 is more secure and faster than the preceding versions since cryptographic weaknesses have been identified over the years for the older TLS versions. \\

\begin{table}
    \centering
    \begin{tabular}{|c|c|c|c|}
    \hline
        \textbf{Category} & \textbf{Frequency}  & \textbf{Percentage (\%)} \\
    \hline
    TLSv1.3 &  107 & 63.3 \\ \hline
    TLSv1.2 & 60 & 35.5 \\ \hline 
    NO TLS & 1 & 0.6 \\ \hline
    Not Accessible & 1 & 0.6 \\ \hline
    \end{tabular}
    \caption{TLS Information}
    \label{tab:tls_information}
\end{table}

Majority of the top websites in the EAC (97.6\%) have valid TLS certificates while the rest either had expired certificates or no form of transport layer encryption in place. The latter exposes sensitive user data to the risk of being intercepted by attackers. Nonetheless, it is worth noting that the cost of obtaining certificates in recent years has been greatly reduced with the introduction of services such as LetsEncrypt.\\ 

\begin{table}
    \centering
    \begin{tabular}{|c|c|c|c|}
    \hline
        \textbf{TLS Certificate State} & \textbf{Frequency}  & \textbf{Percentage (\%)} \\
    \hline
    Valid &  165 & 97.6 \\ \hline
    Expired & 2 & 1.2 \\ \hline 
    No Certificate & 1 & 0.6 \\ \hline
    Not Accessible & 1 & 0.6 \\ \hline
    \end{tabular}
    \caption{State of TLS Certificates}
    \label{tab:state_of_tls_certificates}
\end{table}

\section{Conclusion}
From this research, we were able to identify the extent websites reveal user data to third parties and even unfold the type of leading third-party companies as Analytics and Advertising companies. This user data is potentially accessible to external geopolitical powers. In addition to strict notice and choice mechanisms, countries in the EAC can combat privacy threats by requiring that user data is processed and stored within the EAC. From our results, we found that 35.5\% of the top websites in the EAC use TLSv1.2 whose ciphers have cryptographic weaknesses. Consequently, we recommend the use of TLSv1.3 which is the recommended industry version.\\
We also found that privacy policies of top sites in the EAC are lengthy, difficult to read/understand, and that the extent of third-party tracking is not disclosed on a website’s privacy policy which makes it difficult for users to understand which third parties might be collecting data about them. As a result, website operators should endeavor to not only make their privacy policies shorter but also write them in an easy-to-understand language free from legal jargon. Moreover, they should disclose third parties with whom they share data with through the privacy policies.\\

We also found that privacy policies of top sites in the EAC are lengthy, difficult to read/understand, and that the extent of third-party tracking is not disclosed on a website’s privacy policy which makes it difficult for users to understand which third parties might be collecting data about them. As a result, website operators should endeavor to not only make their privacy policies shorter but also write them in an easy-to-understand language free from legal jargon. Moreover, they should disclose third parties with whom they share data with through the privacy policies. Future work can utilize our results to investigate the extent of compliance of the top websites with data privacy regulations of countries within the EAC.\\

{\small
\bibliographystyle{IEEEtran}
\bibliography{references}

\section*{References}
}
\begin{thebibliography}{00}
\bibitem{b1} Daniel Solove “A Taxonomy of Privacy”, 2006
\bibitem{b2} G. Zhang and K. Yin, “A look at China's draft of Personal Information Protection Law”, 2020. Retrieved from https://iapp.org/news/a/a-look-at-chinas-draft-of-personal-data-protection-law/
\bibitem{b3} E.Goldman,“An Introduction to the California Consumer Privacy Act (CCPA),” Jul.09,2018.https://iapp.org/media/pdf/resource\_center/Intro\_to\_CCPA.pdf(accessed Mar. 10, 2021).

\bibitem{b4} J. R. Mayer and J. C. Mitchell, “Third-Party Web Tracking: Policy and Technology,” in 2012 IEEE Symposium on Security and Privacy, May 2012, pp. 413–427, doi: 10.1109/SP.2012.47
\bibitem{b5} “What are Cookies?,” www.kaspersky.com, Jan. 13, 2021.  https://www.kaspersky.com/resource-center/definitions/cookies (accessed Apr. 15, 2021).
\bibitem{b6} “An overview of HTTP - HTTP | MDN.” https://developer.mozilla.org/en-US/docs/Web/HTTP/Overview (accessed Apr. 15, 2021).
\bibitem{b7} “East African Community.” https://www.eac.int/ (accessed Apr. 15, 2021).
\bibitem{b8} “Data protection,” GOV.UK. https://www.gov.uk/data-protection (accessed Apr. 15, 2021).
\bibitem{b9} T. Libert, “An Automated Approach to Auditing Disclosure of Third-Party Data Collection in Website Privacy Policies”, in Reuters Institute for the Study of Journalism  , 2018.
\bibitem{b10} B. Borena, F. Belanger, and D. Egigu, “Information Privacy Protection Practices in Africa: A Review through the Lens of Critical Social Theory,” IEEE Xplore, J	an. 01, 2015. https://ieeexplore.ieee.org/abstract/document/7070235 (accessed Feb. 23, 2021).
\bibitem{b11} The United Nations Conference and on Trade and Development, “Data Protection and Privacy Legislation Worldwide | UNCTAD.” https://unctad.org/page/data-protection-and-privacy-legislation-worldwide (accessed Apr. 13, 2021).
\bibitem{b12} GSMA, “The State of Mobile Internet Connectivity Report 2020 - Mobile for Development,” Mobile for Development, 2019. https://www.gsma.com/r/somic/ (accessed Apr. 13, 2021).
\bibitem{b13} X. J. Mamakou, D. K. Kardaras, and E. A. Papathanassiou, “Evaluation of websites’ compliance to legal and ethical guidelines: A fuzzy logic–based methodology,” J. Inf. Sci., vol. 44, no. 4, pp. 425–442, 2018.
\bibitem{b14} Information Commissioner’s Office (ICO), “The benefits of data protection laws,” Mar. 12, 2021. 
\bibitem{b15} L. Adedayo, S. Butakov, R. Ruhl, and D. Lindskog, “E-Government web services and security of Personally Identifiable Information in developing nations a case of some Nigerian embassies,” in 8th International Conference for Internet Technology and Secured Transactions (ICITST-2013), Dec. 2013, pp. 623–629, doi: 10.1109/ICITST.2013.6750278.
\bibitem{b16} X. J. Mamakou, D. K. Kardaras, and E. A. Papathanassiou, “Evaluation of websites’ compliance to legal and ethical guidelines: A fuzzy logic–based methodology,” J. Inf. Sci., vol. 44, no. 4, pp. 425–442, 2018.
\bibitem{b17} C. Mutimukwe, E. Kolkowska, and Å. Grönlund, “Information privacy practices in e-government in an African least developing country, Rwanda,” Electron. J. Inf. Syst. Dev. Ctries., vol. 85, no. 2, p. e12074, 2019.
\bibitem{b18} Z. S. Ruhwanya, “Attitudes toward, and awareness of, online privacy and security: a quantitative comparison of East Africa and U.S. internet users,” Thesis, Kansas State University, 2015.
\bibitem{b19} Z. B. Omariba and N. B. Masese, “Security and Privacy of Electronic Banking,” Int. J. Comput. Sci. Issues, vol. 9, no. 3, pp. 432–446, 2012.
\bibitem{b20} C. Mutimukwe, E. Kolkowska, and Å. Grönlund, “Trusting and Adopting E-Government Services in Developing Countries? Privacy Concerns and Practices in Rwanda,” in Electronic Government, Cham, 2017, pp. 324–335, doi: 10.1007/978-3-319-64677-0\_27.

\bibitem{b21} A. Aladeokin, P. Zavarsky, and N. Memon, “Analysis and compliance evaluation of cookies-setting websites with privacy protection laws,” 2017 12th Int. Conf. Digit. Inf. Manag. ICDIM 2017, vol. 2018-January, pp. 121–126, 2017, doi: 10.1109/ICDIM.2017.8244646.

\bibitem{b22} M. A. Bashir, S. Arshad, E. Kirda, W. Robertson, and C. Wilson, “How tracking companies circumvented ad blockers using WebSockets,” in Proceedings of the Internet Measurement Conference 2018, 2018.
\bibitem{b23} P. Papadopoulos, N. Kourtellis, and E. Markatos, “Cookie synchronization: Everything you always wanted to know but were afraid to ask,” in The World Wide Web Conference on - WWW ’19, 2019.
\bibitem{b24} L. A. Abdulrauf, “Giving ‘teeth’ to the African Union towards advancing compliance with data privacy norms,” Information \& Communications Technology Law, vol. 30, no. 2, pp. 1–21, Nov. 2020, doi: 10.1080/13600834.2021.1849953.
\bibitem{b25} Alexa, “Alexa - Top Sites for Countries.” https://www.alexa.com/topsites/countries (accessed Apr. 15, 2021).
\bibitem{b26} T. Libert, timlib/webXray. 2021. Accessed: May 14, 2021. [Online]. Available: https://github.com/timlib/webXray
\bibitem{b27} Alexa, “Alexa - Top Sites in Kenya - Alexa,” www.alexa.com, Apr. 28, 2021. https://www.alexa.com/topsites/countries/KE (accessed Apr. 28, 2021).
\bibitem{b28} Alexa, “Alexa - Top Sites in Tanzania - Alexa,” www.alexa.com, Apr. 28, 2021. https://www.alexa.com/topsites/countries/TZ (accessed Apr. 28, 2021).
\bibitem{b29} Alexa, “Alexa - Top Sites in Uganda - Alexa,” www.alexa.com, Apr. 28, 2021. https://www.alexa.com/topsites/countries/UG (accessed Apr. 28, 2021)
\bibitem{b30} Ugwire, “Top 10 Most Visited Websites In Rwanda 2021 Popular,” UGWIRE, Feb. 25, 2021. https://ugwire.com/top-10-most-visited-websites-rwanda/ (accessed Apr. 28, 2021).
\bibitem{b31} woorank, “Burundi: Websites in Burundi,” index.woorank.com. https://index.woorank.com/en/reviews?countries=BI (accessed Apr. 28, 2021).
\bibitem{b32} Webchart, “Most Popular Websites In Burundi,” Webchart. https://webchart.org/countries/statistics/BI (accessed Apr. 28, 2021).
\bibitem{b33} R. Flesch, “A new readability yardstick.,” Journal of Applied Psychology, vol. 32, no. 3, pp. 221–233, 1948, doi: https://doi.org/10.1037/h0057532.
\bibitem{b34} E. Maris, T.Libert, and J.Henrichsen, “Tracking sex: The implications of widespread  sexual data leakage and tracking on porn websites,” arXiv[cs.CY], 2019.
\bibitem{b35} A. McDonald and L. Cranor, “The Cost of Reading Privacy Policies,” I/S: A Journal of Law and Policy for the Information Society, vol. 4, 2008, Accessed: May. 12, 2020. [Online]. Available: https://lorrie.cranor.org/pubs/readingPolicyCost-authorDraft.pdf

\end{thebibliography}
}

\section*{

\vspace{12pt}

\end{document}